 \documentclass{aastex}

\slugcomment{submitted to \aj}

\shorttitle{C stars in SagDIG and Leo I} 
\shortauthors{Demers \& Battinelli}

\begin{document}


\title{C Star survey of Local Group Dwarf Galaxies. III \\
	The Sagittarius dwarf irregular and \\
	the Leo I dwarf spheroidal galaxies}

\author{Serge Demers\altaffilmark{1}}
\affil{D\'epartement de Physique, Universit\'e de Montr\'eal, Montreal, Qc
H3C 3J7, Canada}
\email{demers@astro.umontreal.ca}

\and

\author{Paolo Battinelli\altaffilmark{1}}
\affil{Osservatorio Astronomico di Roma, viale del Parco Mellini 84, I-00136 Roma, Italy}
\email{battinel@oarhp1.rm.astro.it}

\altaffiltext{1}{Visiting Astronomer, Cerro Tololo Inter-American Observatory.
CTIO is operated by AURA, Inc.\ under contract to the National Science
Foundation.}

\begin{abstract}
We present the latest results of our ongoing homogeneous cool 
C star survey of Local Group dwarf
galaxies. We apply our two color photometric technique to the study of
two small galaxies: the Sagittarius dwarf (SagDIG) and Leo I. We identify
16 C stars in SagDIG and 13  C stars in Leo I. Even though both galaxies
have a known C star population, we identify 7 previously unknown C
stars in each galaxy.
The photometric properties of all the known C stars in each galaxy are
presented. It is shown that our definition of a C star, based on our
photometric criteria, produces a subset of carbon stars with homogeneous
properties useful for population comparison. 
\end{abstract}



\section{INTRODUCTION}

To pursue our ongoing programme (Albert et al. 2000; [Paper I] and 
Battinelli \& Demers 2000, [Paper II])
to determine and compare the photometric properties
of cool C stars in dwarf galaxies, we present the results of a survey of the
Sagittarius dwarf galaxy (SagDIG) and of Leo I, a satellite of the Milky Way. 
This is done to establish
if their mean $M_I$ can be used as a standard candle
and if their mean colors and/or magnitudes are function of the metallicity
or other properties of the parent galaxy. 

SagDIG was discovered independently, on ESO
and SERC survey plates, by Cesarsky et al. (1977) and Longmore et al. (1978).
SagDIG is located in the direction toward the Galactic center 
($\ell = 21^\circ$) at $\alpha_{2000} = 19^h29^m59^s$, $\delta_{2000} =
-17^\circ40'41''$, thus it is at a relatively low Galactic latitude, $b =
-16.3^\circ$.
This dwarf galaxy, located on the outskirts of
the Local Group (van den Bergh 2000), has recently been the subject of two
independent photometric studies. Both investigations determined its distance
from the apparent magnitude of the tip of its giant branch, (TRGB) their
results are consistent with each other. 
Karachentsev, Aparicio \& Makarova (1999)
found a distance of 1.06 $\pm$0.10 Mpc while Lee \& Kim (2000) found 1.18
$\pm$ 0.10 Mpc. Because their adopted color excesses are different, we
detail in 3.1.2 the reasons for our choice of E(R--I). 

Leo I is one of the nine dwarf spheroidal galaxies (dSph) associated with the 
Milky Way. It was discovered, while inspecting Palomar Sky Survey plates,
 by Harrington \& Wilson (1950); it has since that time been the subject
of several investigations.  Leo I is among the most massive dSph satellites of
the Galaxy, it contains a substantial intermediate age population which until fairly
recently was believed to represent also the oldest population of Leo I (Held et al. 2000).
We adopt, for the distance and extinction of Leo I  the 
values obtained by Lee et al. (1993)
from the apparent magnitude of the tip of the giant branch. 
\placetable{tab-1}
Table 1 summarizes the currently known and 
adopted properties of SagDIG and Leo I, $M^T_V$ is the integrated
absolute magnitude.

C stars are known to exist in both galaxies. Cook (1987) surveyed SagDIG
employing a four-filter technique similar to ours. Leo I was 
observed spectroscopically, using a grism technique, by Azzopardi, Lequeux
\& Westerlund (1985;1986) and by Aaronson, Olszewski \& Hodge (1983) 
using near infrared photometry. Our observations of Leo I will be
particularly interesting to link our photometry to a population of
spectroscopically confirmed C stars. As we shall see, our technique
permits only the identification of the coolest C stars,thus defining
a subset with narrower range of properties.

\section{OBSERVATIONS AND DATA REDUCTION}
The observations presented in this paper were obtained at the CTIO 1.5 m
telescope with a 2048$\times$2048 Tek CCD during two runs: a five night 
run in August 1999 and a three night run in April 2001. For the first run,
the telescope was employed at the f/13.5 focus, yielding a pixel size
of $0.24''$ and a field of view of $8.2'\times8.2'$. This field is deemed
suitable for the small galaxies under investigation. It further allows to
evaluate the foreground contamination. For the second run, the f/7.5 focus,
yielding a pixel size of 0.432$''$ and a field of view of $15'\times15'$ was
adopted. To photometrically identify C stars,
we employed the technique used in Paper I and described by Brewer, Richer \&
Crabtree (1995), see also Cook, Aaronson \& Norris (1986). 
Standard Kron-Cousins R and I filters are used along with CN and TiO
interference filters respectively centered at 810 nm and 770 nm. Both filters
have a width of 30 nm. The calibration and data reduction of both runs were
done in the same way. The reader is referred to Paper II of this series 
which detailed the calibration and data reduction procedure.
The journal of observation is presented in Table 2. 
\placetable{tab-2}
Sky flats were obtained each night through each filters. Calibration to the
standard R, I system was done using Landolt's (1992) equatorial standards
observed during the course of the night. 
Extinction coefficients and transformation equations were obtained by
multilinear regressions. Details are presented in Paper II. 

After the standard prereduction of trimming, bias subtraction, and sky
flat-fielding,
the photometric reductions were done by fitting model point-spread
functions (PSFs) using DAOPHOT/ALLSTAR/ALLFRAME series of programs (Stetson
1987, 1994) in the following way: we combine, using MONTAGE2, all the images
of the target irrespective of the filter to produce a deep image devoid of
cosmic rays.
ALLSTAR was then used on this deep image to derive a list
of stellar images and produce a second image where the stars, found in the
first pass, are removed. This subtracted image is also processed through
ALLSTAR to find faint stars missed in the first pass. The second list of
stars is added to the first one. The final list is then used for the analysis
of the individual frames using ALLFRAME. This program fits model PSFs to
stellar objects in all the frames simultaneously.
\section{RESULTS}
\subsection{SagDIG}
\subsubsection{The color-magnitude diagram}
Figure 1 displays the color-magnitude diagram for stars with DAOPHOT computed
errors for $(R-I)$ less than 0.1 mag. It is based on exposures totaling
100 minutes in R and 80 minutes in I. A comparison of this figure with the
CMD produced by Lee \& Kim (2000) shows that the magnitude limit of our
diagram is half a magnitude fainter than the red giant tip of SagDIG,
which is difficult to see because of the numerous foreground stars, many of
them redder than C stars. 
\placefigure{fig1}
\subsubsection{Reddening and adopted distance}
One can estimate, approximately, the reddening toward SagDIG from the
CMD displayed in Figure 1. The $(R-I)$ 
distribution of foreground stars, in the CMD,
is quite sharply limited on the blue side. This limit, corresponding to the
main sequence turnoff of field G dwarfs, 
is seen at $(R-I)$ = 0.39. If we adopt $(R-I)_\circ$
$\approx$ 0.35 for G5 dwarfs (Cox 2000), then the E(R--I) $\approx$ 0.04, 
corresponding
to color excess E(B--V) $\approx$ 0.05. This evaluation confirms the low
reddening estimate by Lee \& Kim (2000), made from a two-color diagram. 
We adopt a low reddening value. G dwarfs, along the line of sight, are
distributed along the first kiloparsecs because, at a distance of 5 kpc 
the line of sight is already 1500 pc below the Galactic disc. 
The position of the blue
ridge in the CMD of NGC 6822 (Letarte, Demers \& Battinelli in preparation)
is located at $(R-I)\approx$ 0.56, a position expected from the published
reddening of that galaxy. Cook (1987) used a similar
technique to evaluate the reddening toward SagDIG. He also concluded that
the reddening is low. 

The adopted
absolute magnitude of SagDIG is based on their respective evaluation of the 
integrated apparent magnitude of this galaxy and again, taking into account 
our adopted reddening. 
\subsubsection{The color-color diagram}
Figure 2 displays the color-color diagram. The same criteria, as adopted in
Paper I,  to define C stars are traced. C stars are stars in the upper box with 
$(R-I) >$ 0.94. The 16 C stars identified in SagDIG are listed in Table 3,
J2000.0 equatorial coordinates are given.
\placefigure{fig2}
\placetable{tab-3}
\subsection{Leo I}
\subsubsection{The color-magnitude diagram}
Leo I being much closer to us that SagDIG, rather short exposures were
secured to reach more than one magnitude below the giant branch tip, 
sufficient to identify C stars. Its color-magnitude diagram is presented
in Fig. 3 The number of foreground stars toward Leo I is much less than in the
direction of SagDIG. nevertheless, a number of very red foreground
stars are seen among the non carbon stars. 

\placefigure{fig3}
\placefigure{fig4}
\subsubsection{The color-color diagram}
The four filter technique allows to easily identify C stars in Leo I. 
There are 13 C stars in Leo I which satisfy our color index criteria, 
they are listed, in Table 4 along with their J2000.0 coordinates. 
As mentioned
above, spectroscopic observations have confirmed at least 19 C stars in
that galaxy. The comparison of the photometric properties of 
our list of C stars with those of the C stars already known, presented
in the next section, shows that results are quite consistent. 
7 C stars, listed in Table 4, are newly identified C stars, including two
of them located $5'$ from its center, a distance well within its tidal
radius (Irwin \& Hatzidimitriou 1995) of 12.6$'$.  
\placetable{tab4}
\section{DISCUSSION}
\subsection{Previously known C stars in SagDIG and Leo I}
Cook (1987) identified 26 C stars in SagDIG, using a photometric technique
similar to ours. Twenty five of his 26 stars were found in our database.
The missing one is near a bright foreground star and it may have been 
rejected because its profile fit did not converge. 
Only 8 C stars are however common to both lists. We present in
Table 5 our magnitude and colors of Cook's C stars. One can see that many
of them have $(R-I)_o < 0.90$ and are thus too blue to be called C stars, 
according to our $(R-I)$ criterion. Star \#4 is redder that
this limit but it is just outside our C star region because its (CN--TiO)
is close to zero. On the other hand, four of our C stars are outside of
Cook's CCD field. Three others, not seen by Cook,  C11, C12, and C13
are among the faintest C stars in our list.

\placetable{tab-5}
There are already known 19 spectroscopically confirmed C stars in 
Leo I (Azzopardi et al. 1985; 1986) plus one (\#20 in Table 6) that is
called by Aaronson \& Mould (1985) a probable C star on the basis of its 
JHK colors and K luminosity. Because our definition of a C star is
restrictive, being limited to stars with $(R-I)_o > 0.90$, 
it is not at all surprising that spectroscopy would have revealed the 
presence of 
warmer/bluer C stars not ``seen'' by our survey. This is indeed the 
case, as one can verify from Table 6, listing our photometric measures
of the 20 C stars compiled by Azzopardi et al. (1986). Only six of them
are red enough to satisfy our color criterion. The reader may ask: why
limit the number of C stars by selecting such a red color? The main
reason for this approach is that we adopt a criterion similar to the
one used by Brewer et al. (1995) and by Nowotny et al. (2001) for their M31 
surveys. By excluding
bluer and fainter C stars, we facilitate their discovery in
external galaxies and we collect a more homogeneous (in I magnitudes)
sample that could eventually be used as standard candles. In fact, one
can easily see, by inspecting Figures 1 and 3, that the bluer C stars
extend to much fainter magnitudes. These bluer stars can be detected
in external galaxies but would require more telescope time. 
\placetable{tab-6}
The numerous foreground Galactic stars toward SagDIG makes the determination
of its C/M ratio highly unreliable. Indeed, we count only 13 C stars, redder than
$(R-I)_o$ = 0.90,  while we count 699 M stars, in the whole $8.4'\times 8.4'$
field, redder than this limit.

Since SagDIG occupies only the central part of the field, counts in
the northern and southern peripheries can be used, in principle, to evaluate
the foreground contribution. These two zones, representing slightly more than
half the field, contain 340 M stars. Thus we estimate that there are 
665$\pm 5\%$ foreground M stars in the field. The uncertainty attached to
the number belonging to SagDIG is such that we cannot quote a meaningful
C/M ratio.

For Leo I it is, however,  easy to evaluate the C/M ratio. To do so, 
we observed a second field,
located $14'$ south east of the center of Leo I. This field, identified as
Leo I comp. in Table 1, contains 41 M stars and no C stars, as expected.
The field centered on Leo I contains 56 M stars. Thus the C/M ratio of 
Leo I is $\approx$ 1. This value is consistent with the C/M ratios of
IC 1613 (0.64) and Pegasus (0.78) and reflects the lower metallicity of
Leo I relative to the two more massive dwarfs.

\section{CONCLUSION}
Leo I and SagDIG are two low mass galaxies of rather different Hubble type 
but of identical absolute magnitude. Few C stars are expected in galaxies of
such low luminosity. More massive dwarfs, like NGC 6822, contain several
hundred C stars. The small number of C stars, seen in these two systems,
provides little else than the mean photometric properties of the C star
population. In more massive dwarfs the spatial distribution of C stars
can be used to map the outer parts of the disc or halo. 
Table 7
summarizes the photometric properties of the C star population in the
two systems under investigation.
\placetable{tab-7}
\placefigure{fig5}
Results, currently on hand, presented in Figure 5, brings further evidences 
that the mean absolute
I magnitude of C stars, $<M_I>$, is nearly constant in galaxies of 
different metallicities and average $\sim -4.7$. The LMC point is based on
the 590 LMC C stars with $(R-I)_o > 0.90$ observed by Costa \& Frogel (1996).
The NGC 6882 data point is from our upcoming publication (Letarte et al.
in preparation). Error bars take into account the uncertainty of the mean
magnitude and the quoted uncertainty of the distance determination. Most
authors quote uncertainty of the true modulus to be $\sim \pm 0.1$ mag.
Freedman (1988) quotes, however, $\pm 0.2$ for her IC 1613 distance
estimate. 
\acknowledgments
This project is supported financially, in part, by the Natural Sciences and
Engineering Research Council of Canada (S. D.).



\clearpage



\figcaption[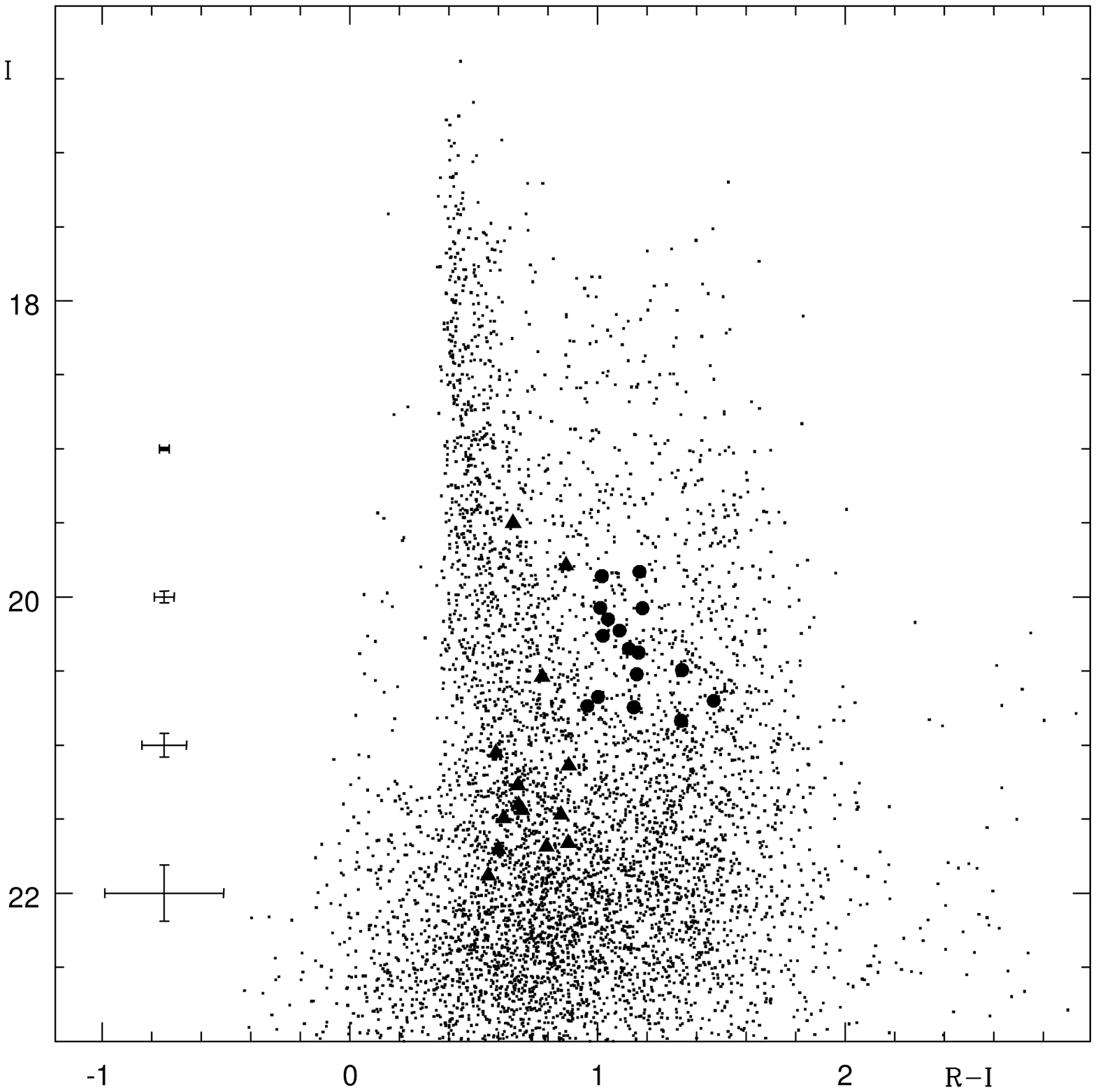]{Color-magnitude diagram of SagDIG. Note the 
well defined 
vertical ridge whose $R-I$ location is function of the reddening along the
line of sight. C stars are shown as big dots. The bluer C stars, identified
by Cook (1987) are represented by triangles.
\label{fig1}}

\figcaption[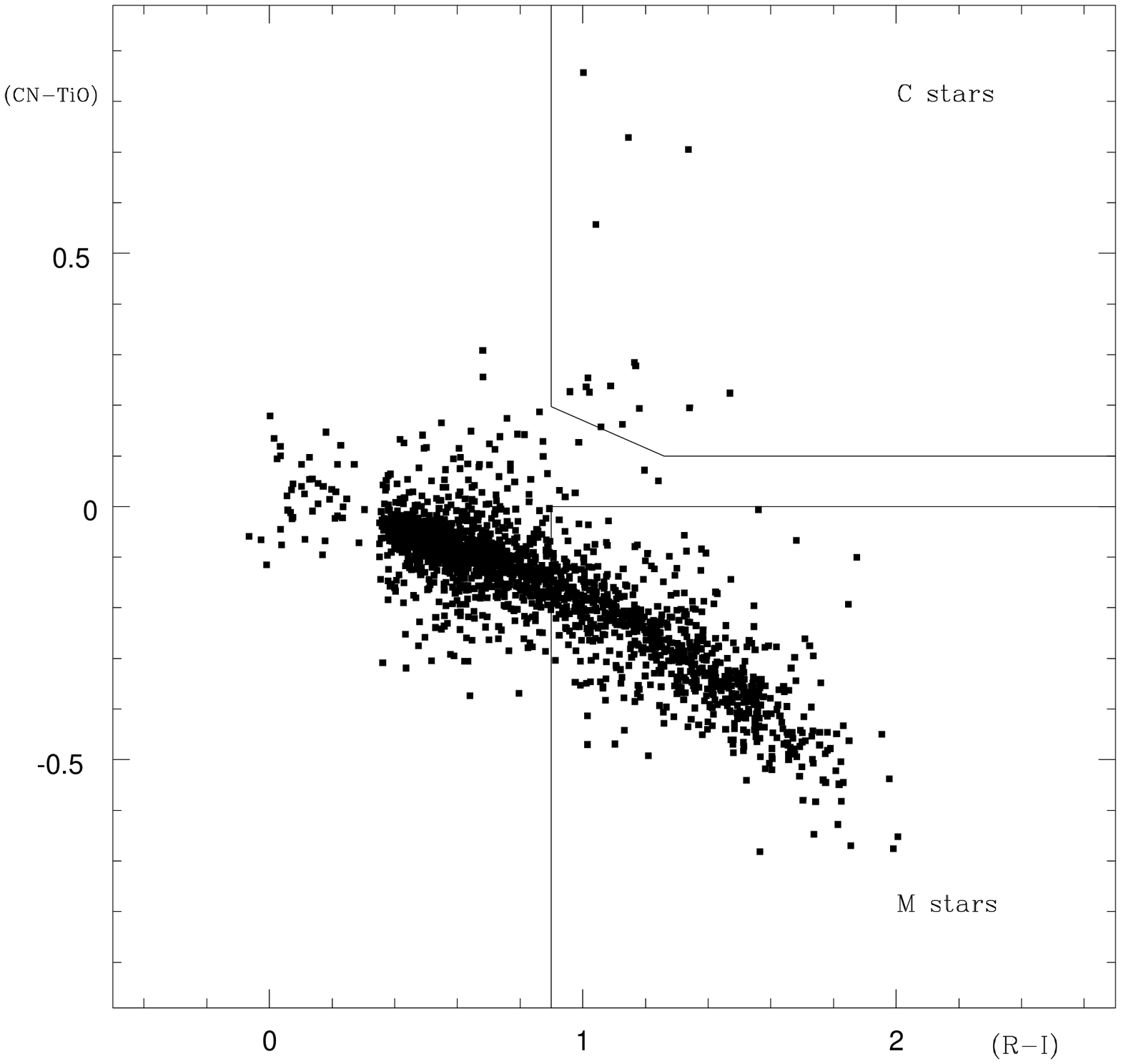]{Color-color diagram of SagDIG.
\label{fig2}}

\figcaption[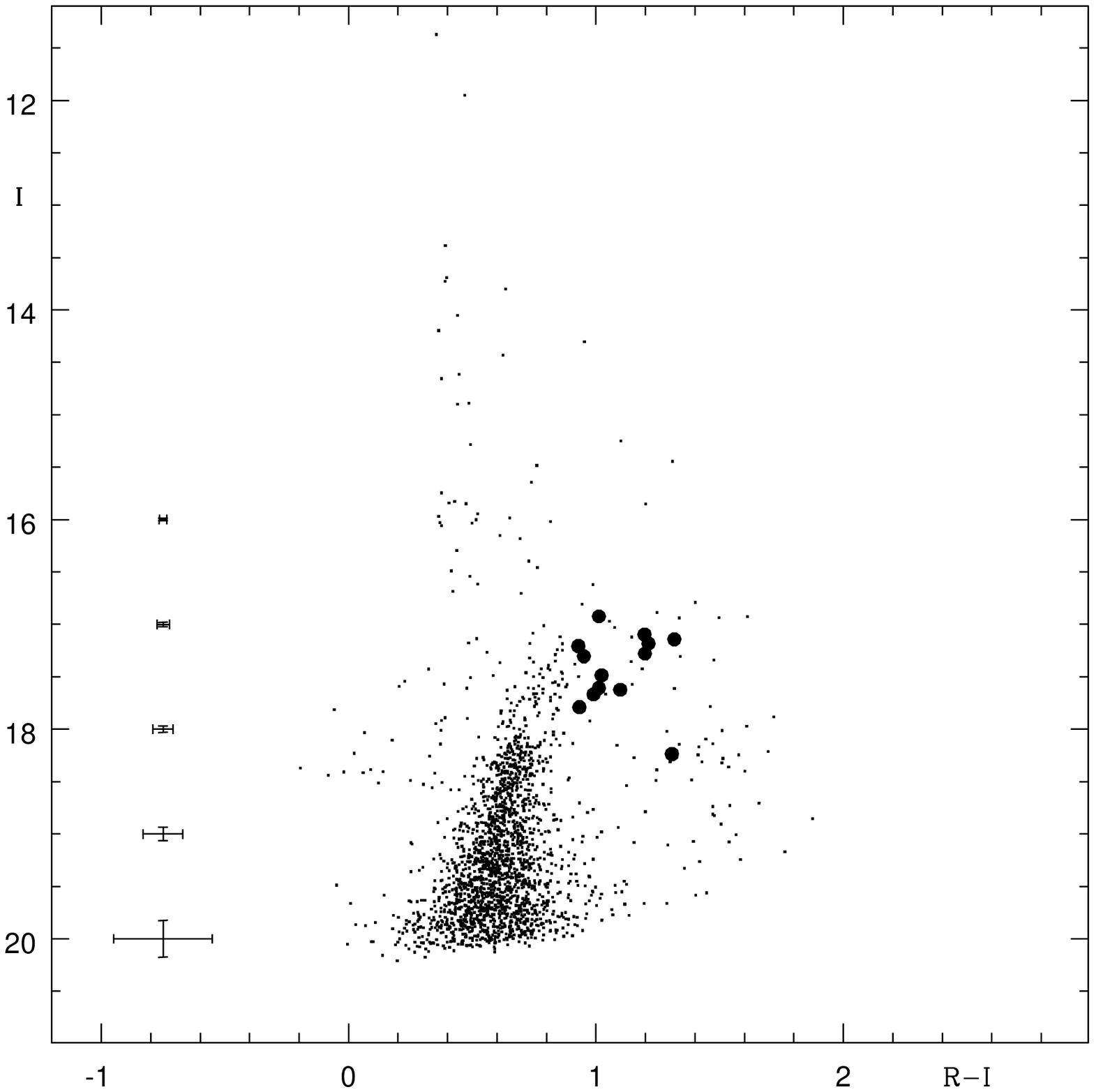]{Color-magnitude diagram of Leo I, C stars are 
shown as big dots while the bluer spectroscopically confirmed C stars
are shown as triangles. \label{fig3}}

\figcaption[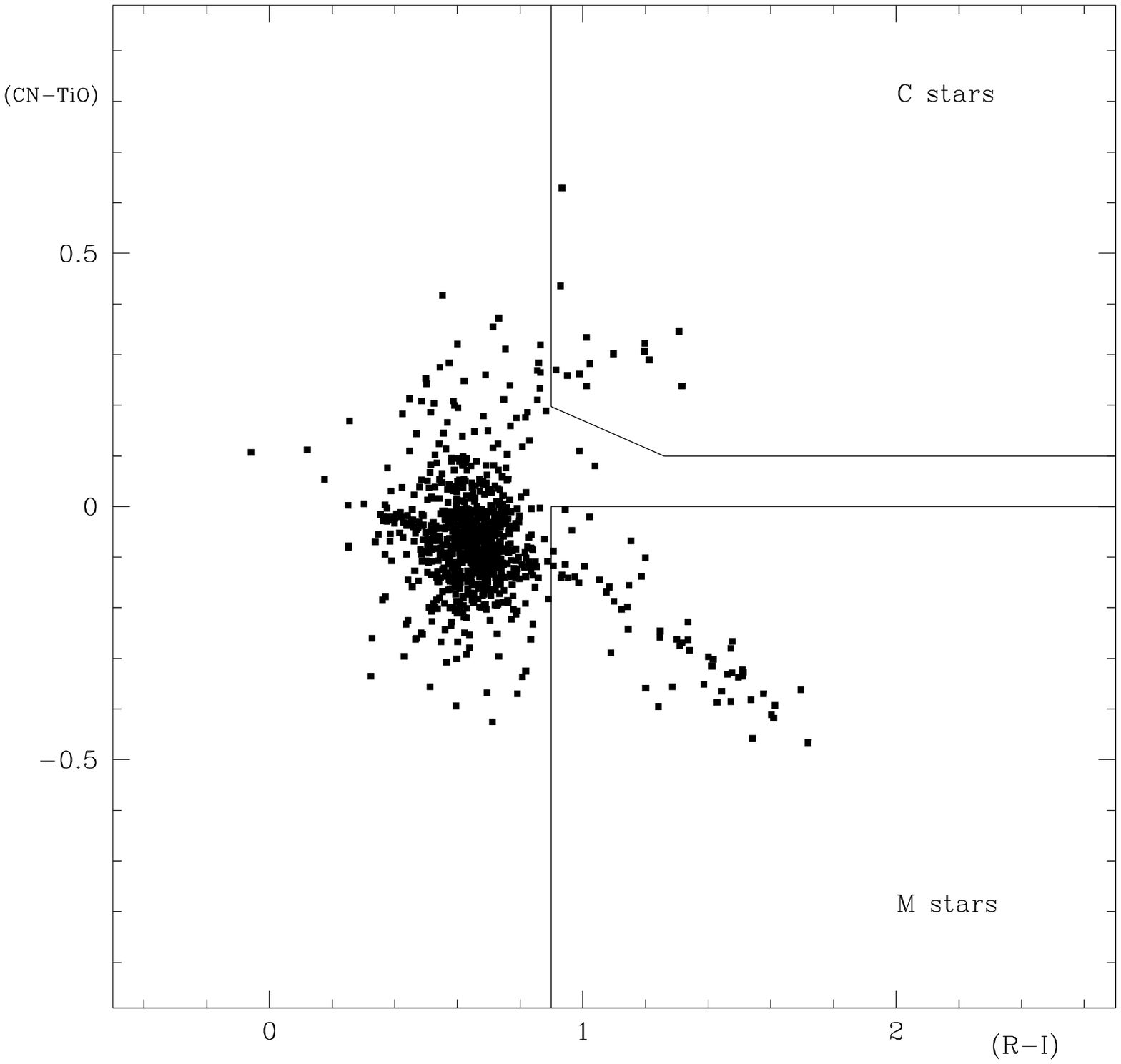]{Color-color diagram of Leo I. 
\label{fig4}}

\figcaption[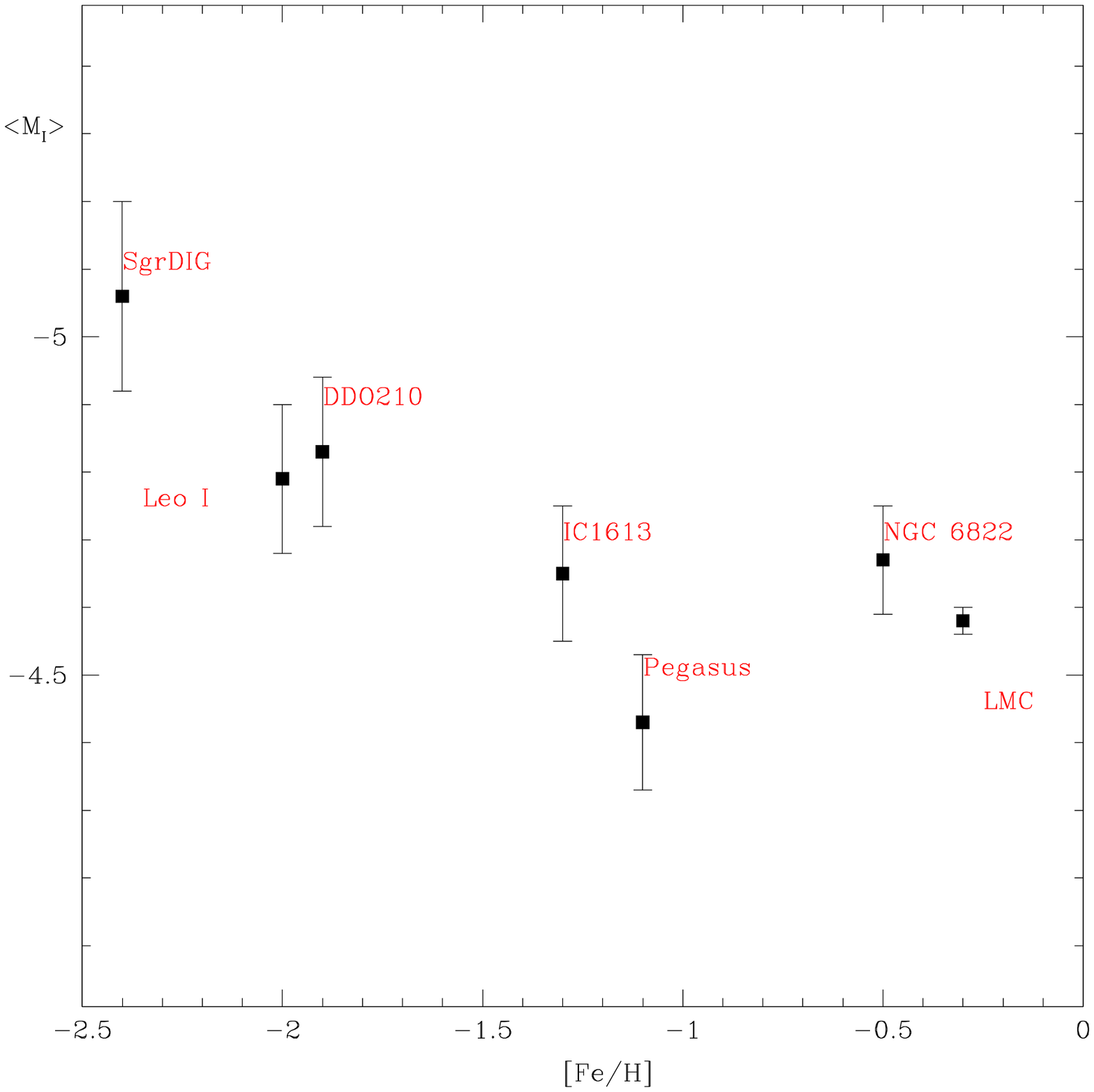]{The mean absolute magnitude, in the I band, 
of C stars, seen in 
different dwarf galaxies, varies little with the metallicity of the parent
galaxy and may be a suitable distance indicator. Error bars reflect mostly
the quoted uncertainties of the distance estimates.
\label{fig5}}



\clearpage

\end{document}